\documentclass[10pt,conference]{IEEEtran}
\usepackage{graphicx}
\usepackage{citesort}
\usepackage{color}
\usepackage[cmex10]{amsmath}
\usepackage{amssymb}
\usepackage{amsfonts}
\usepackage{caption}
\title{Empirical Bayes and Full Bayes\\ for Signal Estimation}
\author{\authorblockN{Yanting Ma, Jin Tan, Nikhil Krishnan, and Dror Baron }
\authorblockA{Department of Electrical and Computer Engineering\\
North Carolina State University\\
Raleigh, NC 27695, USA \\
Email: \{yma7, jtan, nkrishn, barondror\}@ncsu.edu}
\thanks{This work was supported in part by the National Science Foundation under Grant CCF-1217749 and in part by the U.S. Army Research Office under Grant W911NF-04-D-0003. This work was presented at the Information Theory and Application workshop (ITA), San Diego, CA, Feb. 2014.}
}
\IEEEoverridecommandlockouts

\graphicspath{{./} {img/}}
\begin{document}
\maketitle
\newcommand{\xhat}{\widehat{\mathbf{x}}}
\newcommand{\xhati}{\widehat{x}_i}
\newcommand{\note}[1]{ {\bf *** #1 ***} }
\newcommand\numberthis{\addtocounter{equation}{1}\tag{\theequation}}
\begin{abstract}
We consider signals that follow a parametric distribution where the parameter values are unknown. To estimate such signals from noisy measurements in scalar channels, we study the empirical performance of an empirical Bayes (EB) approach and a full Bayes (FB) approach. We then apply EB and FB to solve compressed sensing (CS) signal estimation problems by successively denoising a scalar Gaussian channel within an approximate message passing (AMP) framework. Our numerical results show that FB achieves better performance than EB in scalar channel denoising problems when the signal dimension is small. In the CS setting, the signal dimension must be large enough for AMP to work well; for large signal dimensions, AMP has similar performance with FB and EB.
\end{abstract}
\section{Introduction}
\label{sec.intro}
\subsection{Motivation}
\label{subsec.motiv}
Consider the estimation of an input signal~${\bf x}\in\mathbb{R}^N$ from noise-corrupted measurements~${\bf y=x+z}$, where~${\bf z}\in\mathbb{R}^N$ represents the additive noise.
Ideally, if the statistical characteristics of both the input~${\bf x}$ and the noise~${\bf z}$ are known, then the optimal signal estimator in the mean square error sense can be obtained by the Bayesian method of conditional expectation~\cite{Levy2008}.
In many applications, however, the prior distribution of the input signal may not be available, and thus the conditional expectation cannot be computed.

One of the approaches to resolve unknown priors is based on the {\em minimum description length} (MDL) principle~\cite{Rissanen1984,GrunwaldMDL2010}. The main idea of MDL is that the signal of interest~${\bf x}$ is usually meaningful and compressible, yet the noise~${\bf z}$ is random and incompressible~\cite{DonohoKolmogorov}. Therefore, the signal~${\bf x}$ and noise~${\bf z}$ can be separated by finding the most compressible description of~${\bf x}$ subject to constraints on the noisy measurements~${\bf y}$ and the noise distribution. However, MDL does not always achieve the {\em minimum mean square error} (MMSE)~\cite{DonohoKolmogorov,BaronDuarteAllerton2011}.

Besides the settings where the prior distribution is completely unavailable, there are applications where we may know the prior distribution class while the parameters that characterize the distribution are unknown. For a parametric source that follows an unknown
parameter $\theta$, one can first estimate~$\theta$ based on observed noisy measurements, and then plug the estimated parameter into an appropriate Bayesian estimator. This approach can be categorized as {\em empirical Bayes} (EB)~\cite{Robbins1985}.
However, EB considers only one possible estimate~$\widehat{\theta}$ and discards all other possible estimates that might also be informative. To take into account the uncertainty in $\theta$, a {\em full Bayes} (FB) approach can be applied. In FB, a prior is assigned to $\theta$, thus the posterior of $\theta$, $f(\theta|{\bf y})$, can be computed. The FB estimator is then defined as the weighted sum of the Bayesian estimators with respect to each possible $\theta$, where the weights are the corresponding $f(\theta|{\bf y})$.  

We notice that the FB approach that we discuss in this paper is a mixture over the parameter space. It is worth mentioning that a closely related estimator that mixes over the signal space~\cite{Tan2014XOR} was proposed as {\em universal conditional expectation} (UCE)~\cite{BaronFinland2011}, in which the expectation is computed with respect to a universal prior~\cite{LZ77,Rissanen1984}. It has been proved~\cite{Tan2014XOR} that UCE achieves the MMSE when the input signal ${\bf x}$ is Bernoulli and the noisy measurements ${\bf y}$ are observed from a {\em binary symmetric channel} (BSC)~\cite{Cover06}.

\subsection{Problem setting}
\label{subsec:setting}

{\bf Input distributions: }The input~${\bf x}$ of dimension~$N$ is generated by an independent and identically distributed (i.i.d.) source, and we consider two parametric distributions. Our first distribution is Bernoulli,
\begin{equation}
x_i\sim\text{Bernoulli}(\theta).
\label{eq.Bernoulli}
\end{equation}
That is,~$\mathbb{P}(x_i=1)=\theta=1-\mathbb{P}(x_i=0)$,
where
the subscript~$(\cdot)_i$ denotes the $i$-th component of a vector.
Our second distribution is {\em Bernoulli-Gaussian} (BG),
\begin{equation}
\label{eq.Ber_Gau}
x_i\sim\theta\cdot\mathcal{N}(\mu,\sigma_x^2)+(1-\theta)\cdot\delta(x_i),
\end{equation}
where~$\theta=\mathbb{P}(x_i\neq0)$,~$\mathcal{N}(\mu,\sigma_x^2)$ denotes a Gaussian distribution with mean~$\mu$ and variance $\sigma_x^2$, and~$\delta(\cdot)$ is the delta function~\cite{Papoulis91}. The BG model is often used in sparse signal processing~\cite{GuoWang2007,Starck2010,VilaSchniter2011}.

{\bf Scalar channels: }In scalar channels,
\begin{equation}
\label{eq.scalar}
{\bf y}={\bf x}+{\bf z},
\end{equation}
where~${\bf x},{\bf z}\in\mathbb{R}^N$ are the input signal and the additive noise, respectively.
The noise~${\bf z}$ is i.i.d. Gaussian,~$z_i\sim\mathcal{N}(0,\sigma_z^2)$.
Given the noisy measurements~${\bf y}$, our goal is to find an estimate~$\widehat{\bf x}$ such that~$\mathbb{E}[\|\widehat{\bf x}-{\bf x}\|_2^2|{\bf y}]$ is minimized.

{\bf Matrix channels: }In matrix channels,
\begin{equation}
\label{eq.cs}
{\bf y=Ax+z},
\end{equation}
where~${\bf A}\in\mathbb{R}^{M\times N}$ is the measurement matrix, ${\bf z}$ represents the additive Gaussian noise, and~$z_i\sim\mathcal{N}(0,\sigma_z^2)$.
We assume that~${\bf A}$ is known while the parameters~$\theta$,~$\mu$,~$\sigma_x^2$, and~$\sigma_z^2$ are unknown. Given the measurements~${\bf y}\in\mathbb{R}^M$, our goal is to find an estimate~$\widehat{\bf x}$ such that~$\mathbb{E}[\|\widehat{\bf x}-{\bf x}\|_2^2|{\bf y}]$ is minimized. This channel model~\eqref{eq.cs} covers applications such as compressed sensing (CS)~\cite{CandesRUP,DonohoCS}.

\section{Signal estimation in scalar channels}
\label{sec.scalar}
Let us denote the parameters of the input distributions by a vector~${\bf\Theta}$, i.e., for Bernoulli signals~${\bf\Theta}=(\theta)$ and for BG signals~${\bf\Theta}=(\theta,\mu,\sigma_x^2)$. If all the parameters in~${\bf\Theta}$ are known, then
the {\em Bayesian estimator}, defined as the conditional expectation
\begin{equation}
\label{eq.Bayes}
\widehat{\bf x}^\text{Bayes} =\mathbb{E}[{\bf x}|{\bf y},{\bf\Theta}],
\end{equation}
achieves the MMSE.
In the remainder of the paper, however, we consider settings where the parameters in~${\bf\Theta}$ are unknown.

To perform signal estimation in the EB framework,
we could first perform ML estimation of the parameters~${\bf\Theta}$, and then plug the estimates into the Bayesian estimator~\eqref{eq.Bayes}, we call this estimator the Plug-in:
\begin{equation}
\widehat{\bf x}^\text{Plug-in}=\mathbb{E}[{\bf x}|{\bf y},\widehat{\Theta}_\text{ML}].
\label{eq.plugin}
\end{equation}

Instead of taking the estimated $\widehat{\Theta}_{\text{ML}}$ as the true parameters, the FB approach assigns a prior $f(\Theta)$ to $\Theta$, thus the posterior $f(\Theta|{\bf y})$ can be computed. The model uncertainty is then incorporated by mixing over all possible $\Theta$, we call this mixed estimator a {\em mixture denoiser} (MixD): 
\begin{equation}
\label{eq.MixD}
\widehat{{\bf x}}^\text{MixD} = \int \mathbb{E}[{\bf x}|{\bf y},\Theta]f(\Theta|{\bf y})d\Theta,
\end{equation}
where $f({ {\bf\Theta}|{\bf y}})\propto f( {\bf y}|{\bf\Theta}) f({\bf\Theta})$. Note that the conditional expectation~$E[{\bf x}|{\bf y},{\bf\Theta}]$ in~\eqref{eq.MixD}
is identical to the Bayesian estimator~\eqref{eq.Bayes}
when the prior distribution~$f({\bf x}|{\bf\Theta})$ is available and the true
parameters are ${\bf\Theta}$. Meir and Zhang~\cite{Meir2003} proposed a similar Bayesian mixture framework in machine learning problems, while we apply this mixture approach to signal estimation. If we have some side information about $\Theta$, then an informative prior that captures the side information can be applied to improve the estimation stability. MixD considers the settings when there is no side information about $\Theta$, hence a proper noninformative prior needs to be applied.

We expect that when the prior~$f(\Theta)$ is properly chosen,~$\widehat{\bf x}^\text{MixD}$ can approach the MMSE in scalar channels~\eqref{eq.scalar} as the signal dimension~$N$ grows.
The noninformative prior $f({\bf\Theta})$ is unbiased to any particular ${\bf y}$, and hence does not strongly influence the posterior distribution $f({\bf \Theta|{\bf y}})$. The uniform distribution is an intuitive but ad hoc choice for a noninformative prior. In contrast, the reference prior, introduced by Bernardo~\cite{Bernardo1979}, maximizes the mutual information between the posterior and the prior distribution. For single parameter distributions, the reference prior has been shown to be equivalent to Jeffreys' prior~\cite{Clarke1994}, which is invariant to reparametrization, and hence is a desirable prior distribution.

It is well-known that the ML estimator asymptotically converges to the true parameter almost surely~\cite{Levy2008,Rissanen2012}, and thus the Plug-in signal estimator asymptotically converges to the Bayesian conditional expectation~\eqref{eq.Bayes}.
It has also been verified~\cite{Rissanen2012} that a parameter estimated with a reference prior asymptotically converges to the true parameter asymptotically, and thus it can be conjectured that MixD converges to the Bayesian MMSE.
Recent work by Verd{\'u}~\cite{Verdu2010} has shown that the excess {\em mean square error} (MSE) caused by a mismatch between the true distribution~$f$ and the estimated distribution~$\widehat{f}$ is twice the divergence~$\mathbb{D}(f\|\widehat{f})$ ~\cite{Cover06} between~$f$ and~$\widehat{f}$. In other words, if the divergence~$\mathbb{D}(f\|\widehat{f})$ between the true distribution~$f$ and the estimated distribution~$\widehat{f}$ converges to 0, then the excess MSE vanishes.

Note that MixD~\eqref{eq.MixD} is closely related to another estimator~\cite{Tan2014XOR} that computes UCE with respect to a universal prior for the input signal ${\bf x}$. Consider the estimation for the first entry of~${\bf x}$,
\begin{eqnarray}
\mathbb{E}[x_1|{\bf y}]
=\frac{P(x_1=1,{\bf y})}{P(x_1=1,{\bf y})+P(x_1=0,{\bf y})},
\label{eq:xhat_u}
\end{eqnarray}
where
\begin{eqnarray}
P(x_1, {\bf y})
= \sum \limits_{{\bf x}_2^N \in \{0,1\}^{N-1}} P(x_1\&{\bf x}_2^N) P({\bf y}|{\bf x})\nonumber,
\label{joint_X1_Y}
\end{eqnarray}
${\bf x}_2^N$ denotes the vector~$(x_2,x_3,\ldots,x_N)$, and~$\&$ denotes concatenation. We utilized {\em normalized maximum likelihood} (NML)~\cite{Rissanen2012} as the universal prior of~${\bf x}$ to compute UCE via (\ref{eq:xhat_u}).
When the noisy measurements~${\bf y}$ are corrupted by a BSC, we have verified rigorously~\cite{Tan2014XOR} for Bernoulli inputs that UCE computed via~\eqref{eq:xhat_u} asymptotically achieves the Bayesian MMSE. The main idea in our proof is to show that UCE asymptotically converges to the Plug-in estimator~\eqref{eq.plugin}, which in turn asymptotically converges to the Bayesian estimator and achieves the MMSE. It can be shown that UCE via~\eqref{eq:xhat_u} is closely related to MixD~\eqref{eq.MixD}.
Keeping the rigorous result for the BSC in mind, we believe that UCE with other input and noise distributions asymptotically converges to the Bayesian estimator.

\section{Signal estimation in matrix channels}
\label{sec.AMP_MixD}

We study the matrix channel signal estimation problem in the approximate message passing (AMP)~\cite{DMM2009} framework. AMP can be regarded as an iterative signal estimation algorithm in matrix channels that performs scalar denoising in each iteration. AMP with EB approaches, such as {\em expectation maximization} (EM) and ML, have been studied~\cite{Krzakala2012probabilistic,EMGMTSP,Kamilov2012}. In this section, we first briefly review the AMP algorithm, and then apply MixD as the denoiser within AMP iterations.

\subsection{Review of AMP}
\label{subsec.AMP}

Consider a matrix channel model~\eqref{eq.cs} where the signal distribution follows~${x}_i\sim f_{X}$ and the noise distribution follows ${z}_i\sim f_{Z}$. In the specific model~\eqref{eq.cs} described in Section~\ref{subsec:setting},~$f_{X}$ is Bernoulli~\eqref{eq.Bernoulli} or BG~\eqref{eq.Ber_Gau}, and~$f_Z$ is $\mathcal{N}(0,\sigma_z^2)$.
The measurement matrix~${\bf A}$ has i.i.d. Gaussian entries with unit norm columns on average,
meaning that the matrix entries are~$\mathcal{N}(0,\frac{1}{M})$ distributed and thus the expected value of the column norm is 1.
The AMP algorithm~\cite{DMM2009} proceeds iteratively according to
\begin{align}
{\bf x}^{t+1}&=\eta_t({\bf A}^T{\bf r}^t+{\bf x}^t)\label{eq.AMPiter1},\\
{\bf r}^t&={\bf y}-{\bf Ax}^t+\frac{1}{\delta}{\bf r}^{t-1}
\langle\eta_{t-1}'({\bf A}^T{\bf r}^{t-1}+{\bf x}^{t-1})\rangle\label{eq.AMPiter2},
\end{align}
where~${\bf A}^T$ is the transpose of ${\bf A}$, $\delta=M/N$ represents the measurement rate, $\eta_t(\cdot)$ is a denoising function, and~$\langle{\bf u}\rangle=\frac{1}{N}\sum_{i=1}^N u_i$
for some vector~${\bf u}=(u_1,u_2,\ldots,u_N)$.
In the~$t$-th iteration, we obtain the vectors~${\bf x}^t\in\mathbb{R}^N$ and~${\bf r}^t\in\mathbb{R}^M$. The denoising function~$\eta_t(\cdot)$ is separable, meaning that it is applied component-wise to the noisy measurements, and $\eta_t'({\bf s})=\frac{\partial}{\partial {\bf s}}\eta_t({\bf s})$.
We highlight that the vector~${\bf A}^T{\bf r}^t+{\bf x}^t\in\mathbb{R}^N$ in~\eqref{eq.AMPiter1} can be regarded as noisy measurements of~${\bf x}$ in the~$t$-th iteration with noise variance~$\sigma_t^2$.
The asymptotic performance of AMP can be characterized by a {\em state evolution} (SE) formalism:
\begin{equation}
\sigma^2_{t+1}=\sigma^2_z+\frac{1}{\delta}\mathbb{E}\left[\left( \eta_t\left( X+\sigma^2_{t}W \right)-X \right)^2\right]\label{eq.SE},
\end{equation}
where the random variables~$W\sim\mathcal{N}(0,1)$ and~$X\sim f_{X}$.
Formal statements for SE appear in~\cite{Montanari2012}.
SE (\ref{eq.SE}) implies that in each iteration of AMP, the denoiser estimates ${\bf x}$ from a scalar channel.

A soft-thresholding denoiser is applied in the original derivation of AMP~\cite{DMM2009}, where the optimal threshold of the denoiser in each AMP iteration can be estimated without knowing the input distribution~\cite{DonohoReeves2013,Mousavi2013}.
Donoho et al.~\cite{Donoho2013} generalized the denoising operator to be various minimax denoisers.
With a minimax denoiser, the resulting AMP algorithm is robust to different signal distributions, but may not achieve the Bayesian MMSE.

\subsection{AMP with MixD}
\label{subsec.AMP-MixD-Algo}

We have discussed in Section~\ref{sec.scalar} that MixD is MMSE optimal for the scalar channel~\eqref{eq.scalar}. If we apply MixD as the denoiser~$\eta_t(\cdot)$  in each AMP step~\eqref{eq.AMPiter1}, then \eqref{eq.SE} is MMSE optimal in each iteration. 
Therefore, it can be expected that using MixD as the denoiser in AMP may achieve the MMSE for the matrix channel~\eqref{eq.cs} (we consider the regions where AMP can achieve the MMSE when the exact distribution of the input is known; cases where the MMSE is not achieved are discussed by Krzakala et al.~\cite{Krzakala2012probabilistic} and Zhu and Baron~\cite{ZhuBaronCISS2013}).
In order to make MixD work inside AMP, we need to estimate the effective Gaussian noise in each AMP iteration. The estimated noise variance $\widehat{\sigma}^2_t$ can be calculated as~\cite{Montanari2011}:
\begin{equation}
\widehat{\sigma}^2_t=\frac{1}{M}\sum_{i=1}^M (r^t_i)^2,
\end{equation}
where ${\bf r}^t$ is defined in (\ref{eq.AMPiter2}).
In each iteration of AMP, we replace the denoiser~$\eta_t(\cdot)$ in~\eqref{eq.AMPiter1} by $\eta_{t}^{\text{MixD}}({\bf A}^T{\bf r}^t+{\bf x}^t,\widehat{\sigma}^2_t)$,
which is computed using (\ref{eq.MixD}) with~${\bf y}={\bf A}^T{\bf r}^t+{\bf x}^t$ and~$\sigma_z^2=\widehat{\sigma}^2_t$.

\section{Numerical Results}
\label{sec.numerical}

\subsection{MixD in scalar channels}
\label{subsec.numerical_scalarMUSE}

In this subsection, we compare the MSE of MixD and Plug-in to the Bayesian MMSE in scalar channels~\eqref{eq.scalar}.

{\bf Settings: }In the Bernoulli case, the Bernoulli parameter is $0.05$, and the Gaussian noise is~$\mathcal{N}(0,0.1)$. The input signal dimension~$N$ is evaluated from 10 up to 1,000.
We use Jeffreys' prior for the Bernoulli parameter~$\theta$,
\begin{equation}
f_\text{Jeffreys}(\theta)=\frac{1}{\sqrt{\pi\theta(1-\theta)}}.
\label{eq.Jeffreys}
\end{equation}

In the BG case, the Bernoulli parameter is~$0.1$, the Gaussian part of the signal is~$\mathcal{N}(0,1)$, and the noise is~$\mathcal{N}(0,0.1)$. We use Jeffreys' prior~\eqref{eq.Jeffreys} for the parameter~$\theta$, a uniform prior for~$\mu$,
\begin{equation*}
f(\mu)=
\begin{cases}
\frac{1}{4}\quad\text{if~$\mu\in[-2,2]$}\\
0\quad\text{else}
\end{cases},
\end{equation*}
and a uniform prior for~$\sigma_x$,
\begin{equation*}
f(\sigma_x)=
\begin{cases}
\frac{1}{2}\quad\text{if~$\sigma_x\in[0,2]$}\\
0\quad\text{else}
\end{cases}.
\end{equation*}
Note that we use a uniform prior over the standard deviation~$\sigma_x$ and not the variance~$\sigma_x^2$.
Our current implementation limits the ranges of~$\mu$ and~$\sigma_x$ to~$[-2,2]$ and~$[0,2]$, respectively, and the extension to arbitrary ranges is ongoing work.

\begin{figure}[t]
\begin{center}
\includegraphics[width=85mm]{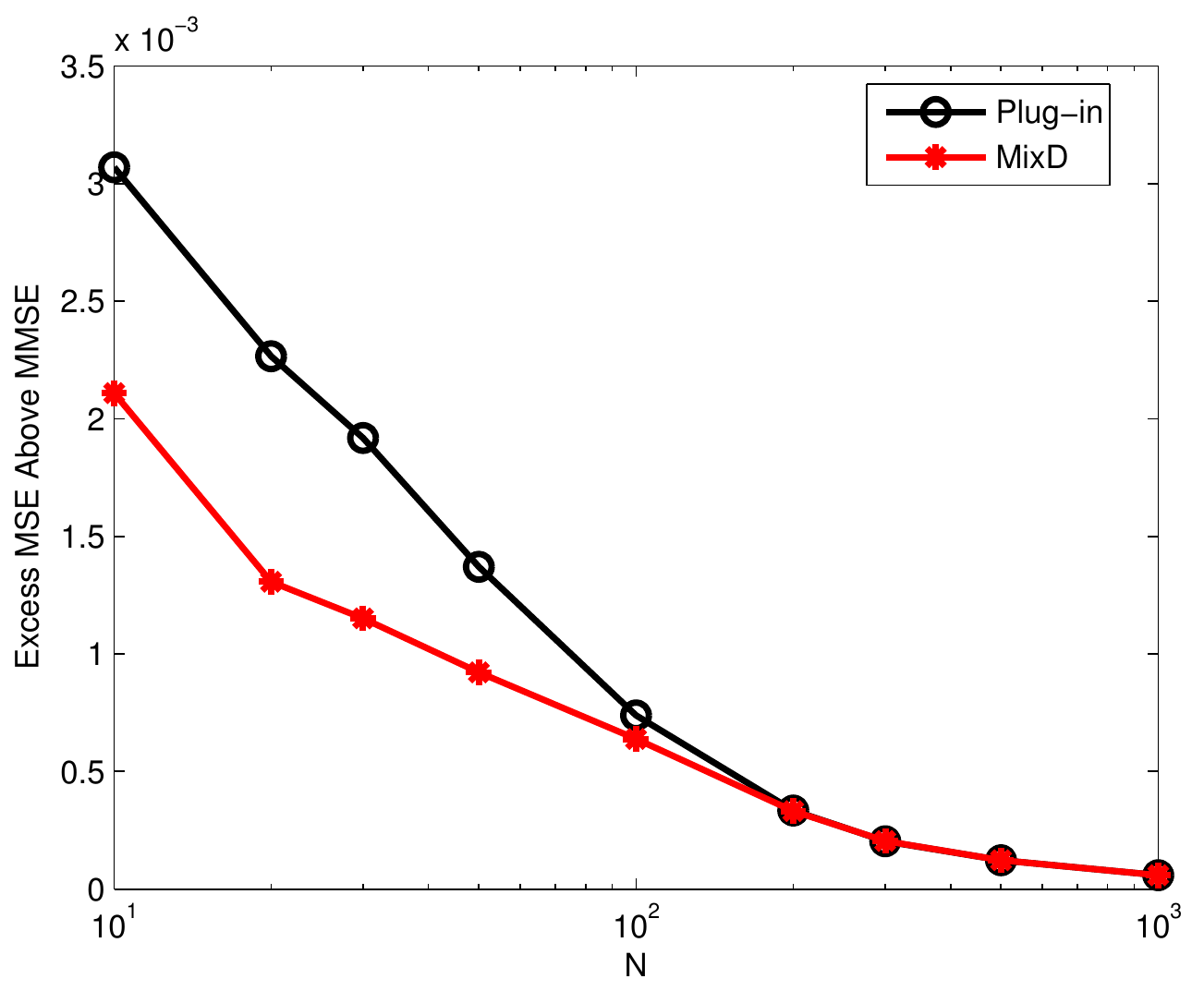}
\end{center}
\caption{{\bf Bernoulli input measured through scalar channel:} We plot the excess MSE above the MMSE achieved by MixD and the Plug-in as functions of~$N$. MixD has better performance for small $N$, and both MixD and Plug-in achieve the MMSE asymptotically. (Signal dimension~$N=10-1,000$, Bernoulli parameter~$\theta=0.05$, and noise variance~$\sigma_z^2=0.1$.)}
\label{fig_scalar}
\end{figure}

\begin{figure}[t]
\vspace*{-2mm}
\begin{center}
\includegraphics[width=82.8mm]{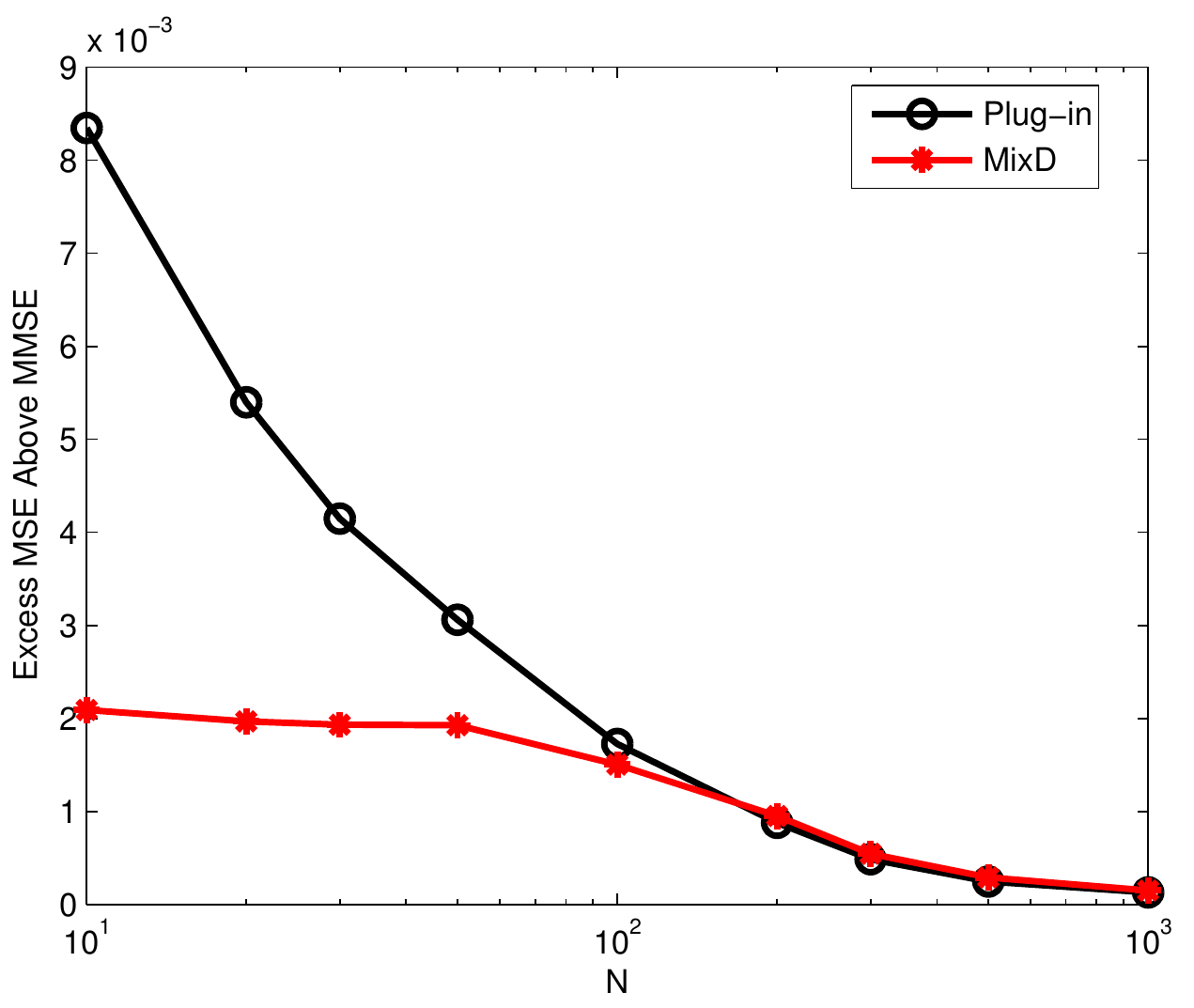}
\end{center}
\caption{{\bf BG input measured through scalar channel:} We plot the excess the MSE above the MMSE achieved by MixD and the Plug-in as functions of~$N$. MixD has better performance for small $N$, and both MixD and Plug-in achieve the MMSE asymptotically. (Signal dimension~$N=10-1,000$, Bernoulli parameter~$\theta=0.1$, Gaussian mean~$\mu=0$, variance~$\sigma_x^2=1$, and noise variance~$\sigma_z^2=0.1$.)}
\label{fig_scalar_BG}
\end{figure}

{\bf Results: }Denote the MSE of MixD and Plug-in by $e_M$ and $e_P$, respectively. We compare the performance of MixD and Plug-in by plotting the excess MSE $(e_M-\text{MMSE})$ and $(e_P-\text{MMSE})$ as functions of~$N$.
Figures~\ref{fig_scalar} and~\ref{fig_scalar_BG} illustrate the results for Bernoulli and BG inputs, respectively. It can be seen from Figures~\ref{fig_scalar} and~\ref{fig_scalar_BG} that MixD achieves lower MSE than the Plug-in when $N$ is comparatively small; MixD performs especially well for BG signals. As $N$ increases, $(e_M-\text{MMSE})$ and $(e_P-\text{MMSE})$ tend to zero, suggesting that both MixD and Plug-in asymptotically achieve the MMSE.

\subsection{AMP-MixD in matrix channels}

\begin{figure}[t]
\vspace*{-1mm}
\begin{center}
\includegraphics[width=85mm]{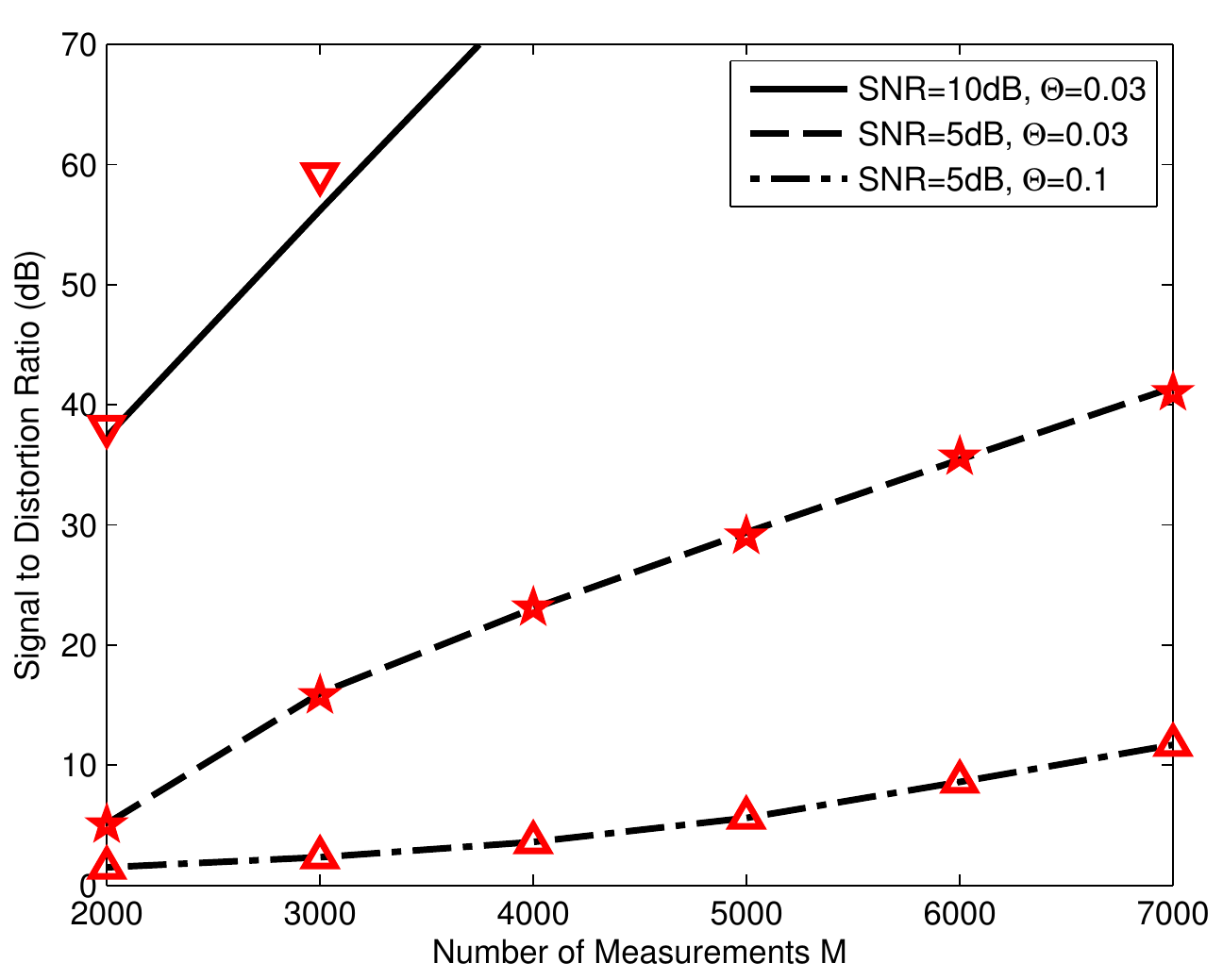}
\end{center}
\caption{{\bf Bernoulli input measured through matrix channel: }We plot the signal to distortion ratio as a function of the
number of measurements $M$ for AMP-MixD.
The curves correspond to the MMSE and the markers
(triangles and stars) represent the MSE performance
of AMP-MixD, which coincides nicely with the
theoretically optimal MMSE. (Signal dimension~$N=10,000$.)
}
\label{fig_cs}
\end{figure}

\begin{figure}[t]
\begin{center}
\includegraphics[width=85mm]{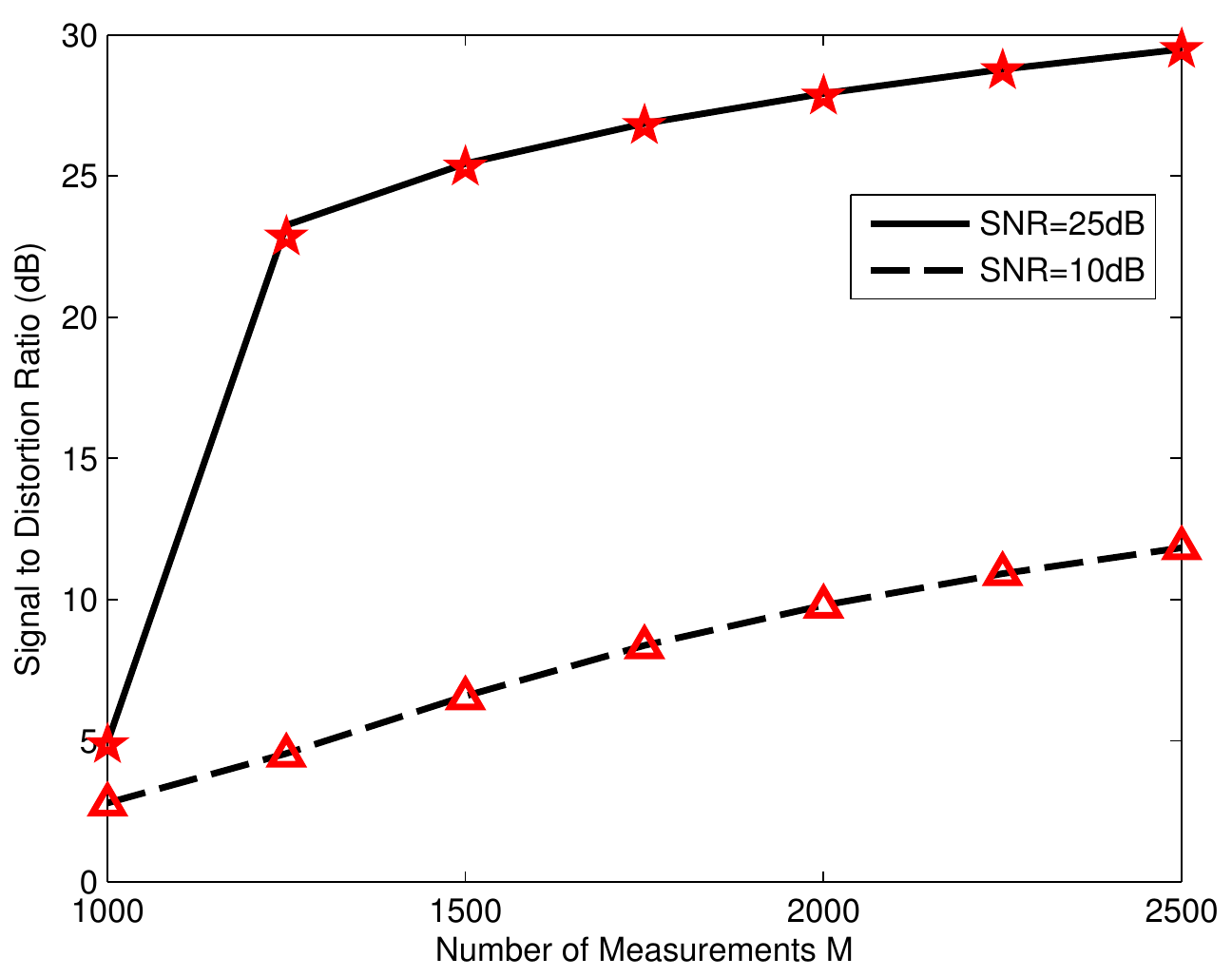}
\end{center}
\caption{{\bf BG input measured through matrix channel:} We plot the signal to distortion ratio as a function of the
number of measurements $M$ for AMP-MixD.
The curves correspond to the MMSE and the markers
(triangles and stars) represent the MSE performance
of AMP-MixD, which coincides nicely with the
theoretically optimal MMSE.
(Signal dimension~$N=5,000$, Bernoulli parameter~$\theta=0.1$, Gaussian mean~$\mu=0$, variance~$\sigma_x^2=1$.)
}
\label{fig_EMBG}
\end{figure}

In this subsection, we evaluate the performance of AMP-MixD for both Bernoulli and BG inputs.

{\bf Settings: }The measurement matrix ${\bf A}$ has i.i.d. Gaussian entries distributed as $\mathcal{N}(0,\frac{1}{M})$. Under this setting, the signal to noise ratio (SNR) of the matrix channel~\eqref{eq.cs} is defined as~$\frac{N\cdot\text{Var}({\bf x})}{M\cdot\text{Var}({{\bf z}})}$, where~$\text{Var}(\cdot)$ denotes variance.

For Bernoulli inputs, the signal dimension $N=10,000$, and the number of measurements $M$ varies from 2,000 to 7,000. The Bernoulli parameter~$\theta=0.03$ or $0.1$, and the SNR is 5 dB or 10 dB.

For BG inputs, the signal dimension $N=5,000$, and the number of measurements $M$ varies from 1,000 to 2,500.
The Bernoulli parameter~$\theta=0.1$, the Gaussian mean~$\mu=0$, the variance~$\sigma_x^2=1$, and the SNR is 10dB or 25 dB.

{\bf Results: }Figures~\ref{fig_cs} and~\ref{fig_EMBG} demonstrate the performance of AMP-MixD for the Bernoulli case and the zero-mean BG case, respectively. The horizontal axis represents the number of measurements~$M$, and the vertical axis represents the signal to distortion ratio, which is defined as the ratio between the signal variance and the MSE of AMP-MixD.
The curves correspond to the theoretically optimal MMSE performance, and the markers
(triangles and stars) represent the performance
of AMP-MixD, which coincides nicely with the
theoretically optimal performance.

Finally, we also simulated the state-of-art algorithm EM-BG~\cite{VilaSchniter2011} and AMP-MixD for the nonzero-mean BG input case, and found that the performance of AMP-MixD and EM-BG are comparable. For brevity, results are not included.

\section{Discussion}
\label{sec.disc}

In this paper, we have shown that the MixD approaches the MMSE
in solving signal estimation problems while adapting to unknown parametric distributions.
Although the results of this paper focus on the Bernoulli and BG parametric
distributions, the concepts can be extended to other distributions, in particular non-i.i.d. signals.
While the Plug-in also approaches the MMSE when the signal dimension~$N$ increases,
readers may notice from Figures~\ref{fig_scalar} and~\ref{fig_scalar_BG} that
MixD approaches the MMSE faster. For example,
in Figure~\ref{fig_scalar} the MSE of MixD is $1.5\cdot 10^{-3}$ above the MMSE
when $N\approx 15$, while the Plug-in achieves the same excess MSE when~$N\approx40$.
In addition to the precision of the signal estimation procedure, another criterion for comparing
algorithms is their speed.
We have noticed that our implementation of the Plug-in runs faster than MixD,
which indicates that the Plug-in could be advantageous in some applications where computational
speed is of paramount importance.
We leave the study of trade-offs between estimation quality and computational requirements for future work.
\section*{Acknowledgments}
We thank Arian Maleki for detailed explanations about his recent work on parameterless AMP~\cite{Mousavi2013};
Phil Schniter for enlightening conversations that significantly influenced this work;
Liyi Dai for inspiring conversations; and Junan Zhu for commenting on the manuscript.
Special thanks to Neri Merhav for providing the original idea for our mixture denoiser.
{
\bibliographystyle{IEEEtran}
\bibliography{cites}
}
\end{document}